\documentstyle[12pt]{article}
\def\be{\begin{equation}\label}
\def\ee{\end{equation}}
\def\bea{\begin{eqnarray}\label}
\def\eea{\end{eqnarray}}
\def\a{\alpha}
\def\b{\beta}
\def\g{\gamma}
\def\d{\delta}

\def\e{\varepsilon}
\def\z{\zeta}
\def\h{\eta}
\def\t{\vartheta}

\begin{document}

\centerline{\large  \bf  Gravitoelectromagnetism
and other decompositions of the Riemann tensor}

\vspace*{1cm}

\centerline{Hans-J\"urgen  Schmidt,
Institut f\"ur Mathematik, Universit\"at  Potsdam}

\smallskip

\centerline{Am Neuen Palais 10,  D-14469 Potsdam, Germany}

\smallskip

\centerline{http://www.physik.fu-berlin.de/\~{}hjschmi, \ 
e-mail: hjschmi@rz.uni-potsdam.de}

\medskip

\begin{abstract}
We present  gravitoelectromagnetism and other decompositions 
of the Riemann tensor from the differential-geometrical  point of view.
\end{abstract}

\bigskip

The word gravitomagnetism is now in common use, even if some
few people (see e.g. page 1306 of Ref. [1]) proposed to write 
gravimagnetism instead. Three of the many recent papers on 
this topic are Refs. [2], [3] and [4]. What is the idea behind this topic ?

\bigskip

The Maxwell field can be represented by the tensor $F_{ij}$, possessing
$4^2 = 16$ real components. By use  of antisymmetry, this figure reduces to 
$6$, but even a six-dimensional space is hard to imagine. However, by 
decomposing it  into two 3-vectors -- the electric and the 
magnetic 3-vector -- the geometric meaning of all the components can be made 
clear, and this is the form where the contact with experimentalists becomes 
possible. Mathematically, these 3-vectors can be obtained by the additional 
introduction of a unit time-like vector field  $u^ i$ and well-known 
transvections with it and with the pseudo-tensor $\epsilon_{ijkl}$.

\bigskip

Now, the analogous procedure shall be made with the gravitational field. 
We have the Riemann tensor $R_{ijkl}$, possessing
$4^4 = 256$ real components. By use  of  the known symmetries, this 
figure reduces to  $20$, but this  twenty-dimensional space is 
even harder to imagine. In the next pages, I will present five different
possibilities  how to arrange  this set of 
components to get a better understandable  system (for details see e.g. Ref. [5]). 

\bigskip

The Riemann tensor $R_{ijkl}$ of  a space-time of dimension $n \ge 3$
 can be decomposed according to several different criteria:

1. The usual one into the Weyl tensor $C_{ijkl}$
plus a term containing  the Ricci tensor $R_{ij}$  plus a term 
containing  the Riemann curvature scalar $R$.

2. Two trace-less parts plus the trace.

3. The  Weyl tensor plus only one additional term. 

4. Two divergence-free parts  plus the trace. 

5. The gravitoelectromagnetic point of view. 

\bigskip

 One should  know, that the mathematical construction 
 of the gravitoelectric field is much elder than the 
word ``gravitoelectric": Einstein calculated the gravitoelectric correction to
the Newtonian potential of the sun - even if Einstein himself never
 had  used  that word. Concerning the invariant characterization of 
gravitomagnetism  one should be careful, because 
 non-flat space-times exist, whose curvature invariants are all 
vanishing, see e.g. Ref. [6]. 

\bigskip

We use  the following two properties of the Riemann tensor
\bea{y1}
R_{ijkl}= - R_{ijlk} \\
R_{ijkl} = R_{klij} \, . \label{y1a}
\eea
The Ricci tensor is the trace of the Riemann tensor:
\be{y2}
R_{ij}= g^{kl}R_{ikjl} \, ,
\ee
 where $g_{kl}$ denotes the metric of the space-time,
and the Riemann curvature scalar is the trace of the Ricci tensor
\be{y3}
R = g^{kl}R_{kl}\, .
\ee
The sign conventions are defined such that in Euclidean signature, the 
curvature scalar of the standard sphere is positive.

For any symmetric tensor $H_{ij}$ we define another tensor
 $H^\ast_{ijkl}$ via
\be{y4}
H^\ast_{ijkl} =H_{ik} g_{jl} + H_{jl} g_{ik}-
 H_{il} g_{jk}- H_{jk} g_{il} \, .
\ee
Then the tensor  $H^\ast_{ijkl}$ automatically fulfils the identities
 eqs. (\ref{y1}) and (\ref{y1a}). For the special case $H_{ij} = g_{ij}$
 we get the simplified form
\be{y4a}
g^\ast_{ijkl} = 2 g_{ik} g_{jl} - 2 g_{il} g_{jk} \, .
\ee

\section{The usual decomposition}

The Weyl tensor $C_{ijkl}$ is the trace-less  part of the Riemann tensor, 
i.e. 
\be{y5}
g^{ik} \, C_{ijkl} = 0\, ;
\ee
it vanishes identically for $n=3$.
Using the notation of eqs. (\ref{y4}) and (\ref{y4a}) we make the ansatz 
\be{y5a}
R_{ijkl}=C_{ijkl} + \a R^\ast_{ijkl} + \b R \,  g^\ast_{ijkl} \, .
\ee
Then the coefficients $\a$ and $\b$ have to be specified such that 
the condition eq. (\ref{y5}) becomes an identity. This
 condition determines the coefficients 
$\a$ and $\b$ uniquely, and  the result is the following: 
\be{y6}
\a = \frac{1}{n-2}\qquad {\rm and}  \qquad    \b =
\frac{-1}{2(n-1)(n-2)}  \, .
\ee
Thus, we get the usual formula
\bea{y6a}
R_{ijkl}=C_{ijkl} + \frac{1}{n-2}
\left( R_{ik} g_{jl} + R_{jl} g_{ik}-
 R_{il} g_{jk}- R_{jk} g_{il}
\right ) \nonumber  \\
 - \,  \frac{1}{(n-1)(n-2)}   R \left(   g_{ik} g_{jl} -  g_{il} g_{jk}
\right)   \, .
\eea

\section{The  decomposition using trace-less parts}

In distinction to the previous subsection, we now perform  a more 
 consequent decomposition into trace and trace-less parts. To this 
end we define $S_{ij}$ as the trace-less part of the Ricci tensor, 
i.e. $g^{ij} S_{ij} =0$ with $S_{ij}= R_{ij} + \kappa  R g_{ij}$
 possessing the unique solution $\kappa = - 1/n$, i.e., 
\be{y7}
S_{ij}= R_{ij} - \frac{1}{n} R \,  g_{ij} \, .
\ee
Then
 the analogous equation to eq. (\ref{y5a}) is 
\be{y8}
R_{ijkl}=C_{ijkl} + \g S^\ast_{ijkl} + \d R \,  g^\ast_{ijkl} \, .
\ee
Again, eq. (\ref{y4}) has been applied. 
Eq. (\ref{y8}) becomes a correct  identity if and only if 
\be{y9}
\g = \frac{1}{n-2}   \qquad {\rm and} \qquad  \d = 
\frac{1}{2n(n-1)} \, .
\ee
So  we get 
\bea{y9a}
R_{ijkl}=C_{ijkl} + \frac{1}{n-2}
\left( S_{ik} g_{jl} + S_{jl} g_{ik}-
 S_{il} g_{jk}- S_{jk} g_{il}
\right ) \nonumber  \\
 + \,  \frac{1}{n (n-1)}   R \left(
  g_{ik} g_{jl} -  g_{il} g_{jk} \right)   \, .
\eea

\section{Decomposition into two parts}
Let us define a tensor 
\be{y10}
L_{ij}= R_{ij} + \z R \,  g_{ij}
\ee
such that a parameter $\e$ exists which makes 
\be{y11}
R_{ijkl}=C_{ijkl} + \e L^\ast_{ijkl}
\ee
becoming a true identity. 
It turns out that this is possible iff 
\be{y12}
\z = \frac{-1}{2(n-1)}   \qquad {\rm and} \qquad  \e = \frac{1}{n-2}\, .
\ee
Thus,  we can write eq. (\ref{y10}) as
\be{y10a}
L_{ij}= R_{ij} -    \frac{1}{2(n-1)}  R \,  g_{ij}
\ee
and eq. (\ref{y11}) as
\be{y11a}
R_{ijkl}=C_{ijkl} +  \frac{1}{n-2}
\left( L_{ik} g_{jl} + L_{jl} g_{ik}-
 L_{il} g_{jk}- L_{jk} g_{il}
\right) \, .
\ee

\section{Decomposition into divergence-free parts}

Now, besides the  identities eqs. (\ref{y1}) and (\ref{y1a}), 
we also use identities  involving  the covariant derivatives, denoted by a semicolon,
 of the Riemann tensor.  The Bianchi identity reads 
\be{y13}
R_{ijkl;m} + R_{ijlm;k} + R_{ijmk;l}=0 \, . 
\ee
Its trace can be obtained by transvection with $g^{ik}$ and reads
\be{y14}
R_{jl;m} + R^i_{\ jlm;i} - R_{jm;l}=0 \, . 
\ee
It should be mentioned, that the transvection with respect to other pairs of 
 indices does not lead to further identities. 
The Einstein $E_{ij}$ tensor is defined as
\be{y15}
E_{ij} = R_{ij} + \lambda \,  R \, g_{ij} \, ,
\ee
where $\lambda$ has to be chosen such that 
the  Einstein tensor is divergence-free, i.e., 
\be{y16}
 E^{i}_{\ j ;i} = 0\, . 
\ee
Using the  trace of eq. (\ref{y14})  (again, there is essentially only one 
such trace), 
\be{y15b}
2R^i_{\ l;i} - R_{;l} =0\, ,
\ee
we uniquely get $\lambda = -1/2$, i.e. the Einstein tensor is 
\be{y15a}
E_{ij} = R_{ij} - \frac{1}{2} R \, g_{ij} \, .
\ee

With the ansatz 
\be{y17}
R_{ijkl} = W_{ijkl} + \h E^\ast_{ijkl} + \t R \,  g^\ast_{ijkl}
\ee
 it holds: 
The coefficients $\h$ and $\t$ are uniquely determined  by the requirements 
that eq. (\ref{y17}) is an identity, and  the divergence of the tensor 
 $ W_{ijkl}$ vanishes: 
\be{y18}
W^{i}_{\ jkl ;i} =0 \, .
\ee
We get the following values of the constants: 
\be{y18a}
\h = 1  \qquad {\rm and} \qquad  \t = \frac{1}{4} \, .
\ee
and then: 
\bea{y15k}
R_{ijkl}=W_{ijkl} + 
 E_{ik} g_{jl} + E_{jl} g_{ik}-
 E_{il} g_{jk}- E_{jk} g_{il}
 \nonumber  \\
 + \,  \frac{1}{2}   R \left(   g_{ik} g_{jl} -  g_{il} g_{jk}
\right)   
\eea
defines a decomposition of the Riemann curvature tensor 
into the divergence-free tensors $W_{ijkl}$, $E_{ij}$, $g_{ij}$
and the scalar $R$. 

\bigskip

It should be mentioned, that for every  $n>2$, the four tensors
$R_{ij}$ eq. (\ref{y2}), $S_{ij}$ eq. (\ref{y7}), $L_{ij}$ eq. (\ref{y10a}) 
 and $E_{ij}$ eq. (\ref{y15a}) represent four different tensors.

It is a remarkable fact, that the coefficients in eqs. (\ref{y15a}) and 
 (\ref{y15k}) do not depend on the dimension $n$. 

\section{The gravitoelectromagnetic point of view}

 The gravitoelectromagnetic point of view  can 
be obtained by the additional 
introduction of a unit time-like vector field  $u^ i$ and well-known 
transvections with it and with the pseudo-tensor $\epsilon_{ijkl}$.

\bigskip

{\Large  {\bf  References} }

\bigskip
\noindent 
[1] H. Urbandtke, Gen. Rel. Grav. {\bf 35} (2003) 1305

\noindent 
[2] V. Faraoni, R. Dumse, Gen. Rel. Grav. {\bf 31} (1999) 91

\noindent 
[3] Jian Qi Shen,  Gen. Rel. Grav. {\bf 34} (2002) 1423

\noindent 
[4]  R. Ruffini, C. Sigismondi (Eds.): Nonlinear 
gravitodynamics - The Lense-Thirring effect, WSPC Singapore (2003)

\noindent 
[5] J. Schouten: Ricci--Calculus, Springerverlag Berlin (1954)

\noindent 
[6] H.-J. Schmidt: Consequences of the noncompactness 
     of the Lorentz group, gr-qc/9512007; 
Int. J. Theor. Phys. {\bf  37} (1998) 691

\end{document}